\documentclass[3p,times,twocolumn]{elsarticle}
 \biboptions{comma,sort&compress}
 
\usepackage{graphicx}
\usepackage{amsmath}
\usepackage{here}
\usepackage{ecrc}

\volume{00}

\firstpage{1}

\journalname{Nuclear and Particle Physics Proceedings}
\runauth{Adriano Di Giacomo}

\jid{nppp}
\jnltitlelogo{Nuclear and Particle Physics Proceedings}

\usepackage{amssymb}

\usepackage[figuresright]{rotating}

\begin{document}

\begin{frontmatter}

\title{Confinement in $QCD$: novelties.
%
} 
 
 \cortext[cor0]{ Talk presented at QCD22,  25th International Conference in QCD (4-7/07/2022,
  Montpellier - FR). }

 \author{Adriano Di Giacomo}

   \address{Universita' di Pisa and INFN Sezione di Pisa, 
   Largo B. Pontecorvo 3.  56127 Pisa  Italy}

\ead{adriano.digiacomo@df.unipi.it}

\pagestyle{myheadings}
\markright{ }
\begin{abstract}
\noindent
We report on recent progress in understanding  confinement of colour in $QCD$ as dual superconductivity of the vacuum. A gauge invariant version of the creation operator of monopoles is
constructed whose vacuum expectation value is the order parameter. This order parameter is gauge-invariant from scratch, has no infrared divergences, is finite in the confined phase and vanishes in the deconfined phase. A natural explanation also emerges of why the electric field  lines in the flux tubes  keep no memory of the colour orientation of the condensing monopoles. A further  by-product  is that the order parameter can be traded with the two-point  vacuum  parallel correlator of the chromo-electric field.
\begin{keyword}  Confinement, Lattice $QCD$, Lattice Quantum Field Theory.


\end{keyword}
\end{abstract}
\end{frontmatter}
\section{Introduction}
 This report  is a completion of  my paper \cite{adg21}. \hspace{.82cm}
 The natural explanation of the vanishingly small upper limits to the existence of quarks in Nature is that they do not exist as free particles (Confinement), due to some symmetry  \cite{D}. The most attractive choice  for that symmetry  is dual superconductivity of the vacuum \cite{'tHP} \cite{m}. 
 
 Confinement is a fundamental problem both in  the standard model and in field theory. Indeed if $QCD$ is the correct theory of the strong sector the mechanism has to be built in $QCD$ at large distances.
 
 The physical idea is as follows.
 In an ordinary superconductor electric charge conservation is spontaneously broken. The ground state is a superposition of states with different number of basic units of charge [ Cooper-pairs]. If $C$ is the creation operator of a Cooper pair  $\langle C \rangle$ is the order parameter: $\langle C \rangle=0 $ in the normal state, $\langle C \rangle \neq 0$ in the superconducting state.
 There magnetic field is squeezed into Abrikosov flux tubes with constant Energy/Unit-Length, and thus magnetic charges are confined. 
 
  In a dual superconductor electric and magnetic are interchanged: $\langle C \rangle$ is replaced by $\langle \mu \rangle $ with $\mu$ the creation operator of a monopole. In the phase $\langle \mu \rangle \neq 0$ there is dual superconductivity,
 electric field is squeezed into dual-Abrikosov flux tubes with constant Energy/Unit-Length and electric charges are confined. In the phase $\langle \mu \rangle =0$ the system is normal and there  is no confinement.
 These features of $QCD$ can be studied on the lattice.
 
 In the same way as the conjugate momentum $p$  generates translations of the position x of a particle $ \exp( i pa)  | x \rangle = | x+a \rangle$, a monopole is created  by shifting the appropriate colour component of the gauge field by  the classical 
 monopole configuration \hspace{.01cm}$ \frac{1}{g}\vec A^{Cl}_{\perp} $ in the transverse gauge \hspace{.01cm} \cite{Dig}\cite{DP} \cite{ddpp}.
 
 \begin{equation}
 \mu (\vec x,t) =  \exp( i \frac{1}{g}\int d^3 y\vec  A^{Cl}_{\perp} (\vec x -\vec y) \vec E_{\perp} (\vec y,t) ) \label{mu}
 \end{equation}
 \begin{equation}
 \mu (\vec x,t) | A_{\mu}(\vec x,t)\rangle =  | A_{\mu}(\vec x,t) +  \frac{1}{g} \vec  A^{Cl}_{\perp} (\vec x -\vec y)\rangle \label{mup}
\end{equation}

$g$ is the gauge coupling and the factor $\frac{1}{g}$ comes from Dirac relation between electric and magnetic charge and  is general independent on the specific gauge group \cite{Dirac}\cite{tH}\cite{Poly}. 
Eq's(\ref{mu}) (\ref{mup}) are strictly speaking  written for $U(1)$ gauge theory where a single gauge field is present. In that case $\mu$ Eq(\ref{mu})  is invariant under gauge transformations   both of the classical external field and of the quantum field, and with it the order parameter $\langle \mu \rangle$. For non-abelian gauge groups $\vec A_{\perp} $ and its conjugate momentum $\vec E_{\perp}$ are replaced by the components in colour space appropriate to the $SU(2)$ sub-group in which the monopoles live (Abelian projection \cite{'tHoo}). 

\hspace{.01cm} $\langle \mu \rangle = \frac{1}{Z(S)} \int [dU] \mu \exp (-\beta S)$ \hspace{.2cm}with $\beta =\frac{2N}{g^2}$ and $S \propto \vec E_L^2 + \vec H_L^2 $ the action. It is clear from
the action that the canonical field $\vec E_{\perp} (\vec y,t) \propto \frac{1}{g} (E_L)_{\perp}$ so that  $\mu$ has the form $\mu =\exp(-\beta \Delta S)$  or
\begin{equation} 
\hspace{1.5cm}\langle \mu \rangle =\frac{Z(S+\Delta S)}{Z(S)} \label{frac}
\end{equation}
Eq(\ref{frac}) is valid independent of the gauge group. This suggests a way to compute $\langle \mu \rangle $ on the lattice \cite{DP} \cite{ddpp} by defining $\rho(\beta) \equiv \frac{\partial \ln (\langle \mu(\beta) \rangle)}{\partial \beta}  = \langle S \rangle_S- \langle S+\Delta S \rangle _{S + \Delta S}$ 
The label on the right indicates the action used to perform the average. $\rho$ is an easy quantity to compute numerically on the lattice, and from it $\langle \mu \rangle$
\begin{equation}
\langle \mu(\beta) \rangle =\exp(\int_0^{\beta} \rho(\beta') d \beta') \label{romu}
\end{equation}
In fact $\rho(\beta)$ is computed on a finite lattice: in the thermodynamic limit $V \to \infty$ it should remain finite in the confined phase, making $\langle \mu \rangle \neq 0$ and should diverge negative  in the deconfined phase making $\langle \mu \rangle = 0$. This can not be proved numerically, but only by  use of an analytic argument. 

 We will do  that by use of of the expansion of $\rho$ in powers of $\Delta S$ discussed in Appendix A  of \cite{adg21}

 \begin{equation}
 \rho= - \sum_{0}^{\infty} \frac{(- \beta)^{n}}{n!}\langle \langle \Delta S^{n+1}\rangle \rangle -\langle \langle S \sum_{1}^{\infty} \frac{(- \beta )^{{n}}}{n!}  \Delta S^n \rangle \rangle \hspace{.3cm} \label{app}
 \end{equation}
 
 $\langle \langle  .. \rangle \rangle  \equiv$ connected part of $\langle ..\rangle_S $
 The result will be that the order parameter is well defined for $U(1)$ gauge theory, where confinement  is indeed produced by dual superconductivity of the ground state. For higher groups the quantity $\rho$ appearing in Eq(\ref{romu}) diverges  as $V^{\frac{1}{3}}$ both in the confined and in the deconfined phase, whenever the monopole lives in a subgroup transforming locally  under gauge transformations (local Abelian Projection\cite{'tHoo}) . The deep reason is that creating a monopole breaks an $SU(2)$ symmetry by the $vev$ of the Higgs field. If that symmetry is local it cannot be broken in a gauge invariant way \cite{elitzur}.The only way out is that the $SU(2)$ subgroup in which the monopole lives be gauge invariant. This will uniquely indicate  the correct construction of the order parameter.
 
\section{Computing $\rho$ analytically.}
On lattice $S= \sum_{n, \mu \nu}\Re[ P_{\mu \nu}-1]$ with $P_{\mu \nu}$ the plaquette

\hspace{.1cm}$ P_{\mu \nu}(n)=\frac{1}{N}Tr[ U_{\mu}(n) U_{\nu}(n+\hat \mu) U^{\dagger}_{\mu}(n+ \hat \nu)U_{\nu}^{\dagger}(n)]$

and\cite{dlmp}
\hspace{.5cm}$S+\Delta S= \sum_{n, \mu \nu}\Re [ P'_{\mu \nu}(n)-1]$ .

$P'_{\mu \nu}(n)= P_{\mu \nu}(n)$ for all $n, \mu, \nu$ except  $\mu=i,\nu=0$ at the time   at which  the monopole is created  which we shall conventionally choose as $t=0$ by use of time translation invariance.There

$P'_{\mu \nu}(\vec n, 0) = \frac{1}{N} Tr[U_i(\vec n, 0) U_0( \vec n + \hat i,0)M_i (\vec n + \hat i)U_i^{\dagger}	(\vec n, 0+1)U_0^{\dagger}(\vec n, 0)] $

 $M_i (\vec n) = \exp({ig}T_3  {A^{Cl}_{\perp}}_i (\vec n - \vec x) ) $ \cite{dlmp}. 
 
 $\vec x$ is the position of the monopole and $T_3$ the third generator of the $SU(2)$ sub-group in which the monopole lives. A local abelian projection has been assumed. 
 
 For $U(1)$  gauge theory  $T_3 $ is replaced by $1.$

 The quantity $\Delta S$ is immediately computed\cite{adg21}. It is non zero only on the hyperplane $t=0$.
 \begin{equation}
 \Delta S = \sum_{\vec n   i}\big ([C_i(\vec n)-1]\Re P_{i0}(\vec n,0)-S_i(\vec n)\Im Q_{i0}(\vec n,0)\big )\hspace{.2cm}\label{deltas}
 \end{equation}
$ C_i (\vec n) \equiv \cos(\frac{g}{2} A^{Cl}_i(\vec n + \hat i -\vec n))$,
$S_i(\vec n) \equiv \sin(\frac{g}{2} A^{Cl}_i(\vec n + \hat i -\vec n))$
\begin{eqnarray}
Q_{i 0}(\vec n, 0)~=~\frac{1}{N}Tr[U_i(\vec n,0) U_0(\vec n + \hat i,0)T_3 \nonumber \\U_i^{\dagger}(\vec n, 1)U_0^{\dagger}(\vec n, 0)] \label{qio}
\end{eqnarray}
For $U(1)$ $T_3 \to1$ and $Q_{i0}= P_{i0}$. 
Eq(\ref{qio}) can also be read as $Q_{i0}(\vec n, 0)= Tr [ G_{i0} T_3]$
\begin{equation}
G_{i0} \equiv \frac{1}{N} U_i^{\dagger}(\vec n, 1)U_0^{\dagger}(\vec n, 0)U_i(\vec n, 0) U_0( \vec n +~ \hat i,0) \label{ef}
\end{equation}
Since the classical field $A^{Cl}_i(\vec n )$ decreases as $\frac{1}{n}$ at large distances 
\begin{equation}
C_i(\vec n) - 1\propto  \frac{1}{n^2} \hspace{.4cm} S_i(\vec n)\propto \frac{1}{n}  \hspace{.6cm} n \to \infty \label{asY}
\end{equation}
Eq(\ref{app}) gives $\rho$ as a sum on the positions of the vacuum correlation functions of products of factors $[\Re P_{i0} (\vec n )(1- C_i(\vec n))]$  and $[S_i(\vec n) \Im Q_{i0}(\vec n)]$. 
The latter always appears an even number of times since it is odd under charge conjugation. In the confined phase there is a mass-gap, the correlations fall down exponentially  and all the integrals on relative distances are convergent.
A dependence  on the sum of coordinates however is left in the factors $(1- C_i(\vec n))$ and
$S_i(\vec n)$. Due to their behaviour at large distance Eq(\ref{asY}), divergences as $V^{\frac{1}{3}}$ in the thermodynamic limit  only exist in the terms proportional to $\Re P_{i0} (\vec n )(1- C_i(\vec n))$ and $[S_i(\vec n) \Im Q_{i0}(\vec n)]^2$. These we will call Kinematic divergences.We shall compute them and prove their exact cancellation  in the $U(1)$ gauge theory in the next section. We will also show that for higher groups, whenever the orientation of the monopole in colour space is not gauge invariant, the divergence does not cancel neither in the confined nor in the deconfined phase and the order parameter does not exist.

In the deconfined phase there exists no intrinsic scale, everything depends only on the relative coordinates  and $\rho$ is expected to behave as
\begin{equation}
\hspace{1.cm}\rho \propto_{V\to \infty}   -K\ln V \hspace{1.cm} K > 0 \label{decon}
\end{equation}
One can trace the origin of that behaviour in the expansion  Eq(\ref{app}) \cite{adg21}. Each factor proportional to $P_{i0}(\vec n)$  enters with a scale dimension $-3$ 
\begin{equation}
d^3n (1- C_i(\vec n))\Re P_{i 0} (\vec n)\approx \frac{d^3n}{n^2}a^4 \vec G_{i 0}\vec G_{i 0}\hspace{.2cm}  [dim -3]
\end{equation}
 Each factor proportional to $Q_{i0}(\vec n)$ enters as
 \begin{equation}
d^3n S_i(\vec n)\Im Q_{i0}(\vec n)\hspace{1.cm} \approx  \frac{d^3n}{n }a^2 \vec G_{i 0}\hspace{.55cm} [dim \hspace{.1cm} 0]
\end{equation}
$a \equiv $  lattice spacing.
The behaviour of the form Eq(\ref{decon}) can only come from terms in Eq(\ref{app}) containing only factors $Q_{i0}$. In particular it is proved \cite{adg21} that, after removal of the divergent part, the term quadratic in $Q_{i0}$ is negative definite  in agreement with Eq(\ref{decon}).
\section{Computing  the kinematic divergences. $U(1)$ theory.}
As anticipated above the divergent part of $\rho$ comes from the lowest terms of the expansion Eq(\ref{app})

\hspace{.2cm}$ \rho_{div} =  \langle \langle - \Delta S + \beta S \Delta S+ \sum_{i, \vec n_1, \vec n_2}  [ \beta \Im Q_{0i}(\vec n_1) \Im Q_{0i}(\vec n_2	)  $

\vspace{.3cm}

$-\frac{1}{2}\beta ^2  S \Im Q_{0i}(\vec n_1 ) \Im Q_{0i}(\vec n_2) ] S_i^2(\vec n) \rangle \rangle  $,
\hspace{.4cm}$\vec n=\frac{\vec n_1 + \vec n_2}{2}$ \hspace{.3cm}

\vspace{.3cm}
For any gauge group this gives, in a strong coupling expansion,              \cite{adg21}
\vspace{.2cm}

$\rho_{div} = \sum_{k=0}^{\infty}\frac{\beta^{2k +1}}{2k !}(k+1) \rho_{div} (k)$ \hspace{.5cm}with

\begin{eqnarray}
\rho_{div} (k)=
\sum_{ i ,\vec n_1 ,\vec n_2}\langle \langle [\Re P_{i0}(\vec n_1) \Re P_{i0}(\vec n_2) \nonumber \\
- \Im Q_{i0} (\vec n_1)\Im Q_{i0}(\vec n_2)] S^{2k}\rangle \rangle  \label{ff}
\end{eqnarray}
Fig. 1 shows the lowest non trivial contribution  ( order  $\beta ^5$ ) to $\rho_{div}$. For gauge group $U(1)$ $Q_{i0} = P_{i0}$ the dots are equal to 1 and the two cubes cancel. It is easy to prove that the two terms in Eq(\ref{ff}) cancel each other exactly to all orders, or that  $\rho_{div}=0$. The reason for that is a symmetry and will be discussed below.

 \begin{figure}[ht]
   \begin{center}
\includegraphics[width=.5\textwidth,clip=]{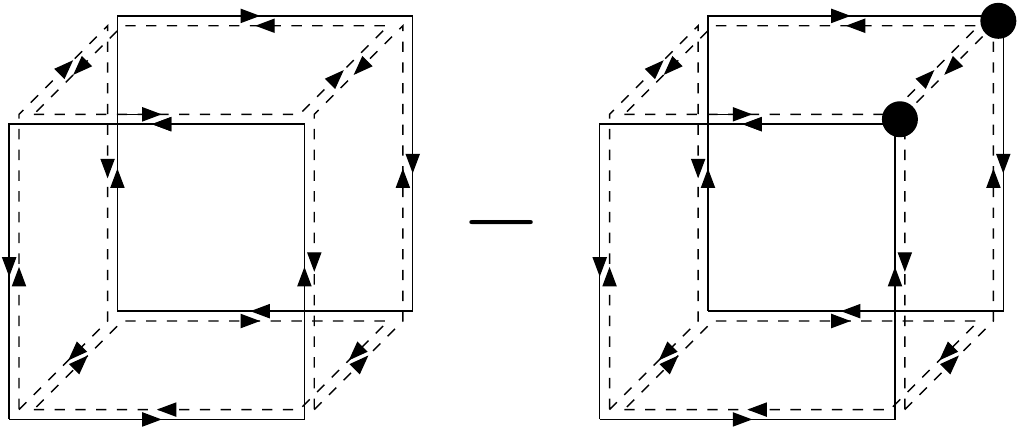}
 \caption{  $\beta^5$ contribution $\rho_{div}$.
Dots denote $T_3$ insertions in the solid lines.}
\end{center}
\end{figure}

For non abelian gauge groups, say $SU(N)$, the second cube vanishes as seen by direct computation, whilst the first one  $=\frac{1}{N^4}$: they do not cancel each other so that the order parameter does not exist. $\rho$ diverges linearly at large sizes both in the confined and in the deconfined phase and $\langle \mu \rangle =0$ or  $\langle \mu \rangle =+ \infty$.
  
Indications of that were found numerically \cite{ddd} \cite{bcdd}. It is obvious that the two cubes will never cancel: the first one is gauge-invariant, the second one is not as long as the $T_3$ insertions are  gauge dependent.

The symmetry which makes $\rho_{div}=0$ for gauge group $U(1)$ is the absence of electric charges  $\partial_i  F_{0i} =0$\cite{adg21}. Indeed expressing the quantity in Eq(\ref{ff}) in terms $P_{i 0)}(\vec n)$ and  $P_{i 0}^*(\vec n)$ the mixed terms cancel (cancellation of the two cubes in Fig.1) and a term $\frac{1}{2}\langle \langle P_{i0}(\vec n_1) P_{i0}(\vec n_2) + c.c.\rangle \rangle]$ is left which is not even displayed in Fig. 1. There is indeed no way of constructing a cube in which the two external plaquettes have the same orientation: this would give a net flux of the electric field across the closed surface of the graph which is not allowed.

For the $U(1)$ theory it was known analytically that monopoles condense below a certain value of $\beta =\frac{2}{g^2}$, $\beta_c$\cite{FM} \cite{PC}. In \cite{DP} it was shown that their definition of $\mu$ is equivalent
to that of Eq(\ref{mu}). Lattice simulations give a determination of $\beta_c$ and evidence that for $\beta \le \beta_c$ electric charges are confined, i.e. the Wilson loops obey the area law. In $U(1)$ theory electric charges are confined by dual superconductivity of the vacuum.
\section{A gauge-invariant order parameter for  $QCD$.}
 For non-abelian gauge groups we have shown that the kinematic divergences do not cancel and as a consequence the order parameter does not exist, whenever the monopoles live in a colour  $SU(2)$ which is not gauge invariant.  A way out is  to replace $T_3$ in Eq(\ref{qio}) by its parallel transport to a point on the surface say at spatial infinity. \hspace{.3cm} $T_3 \to \bar T_3$ \hspace{.3cm} 
 \begin{equation}
 \bar T_3  =  V_C(\vec n, \infty)\hspace{.1cm} T_3 \hspace{.1cm} {V^{\dagger}}_C (\vec n, \infty) \label{transp}
 \end{equation}
 $V_C $   is the parallel transport from $\vec n$ to $\infty$ along a line $C$. The transformation is a rotation of $T_3$ and does not affect the properties of the order parameter \cite{adg21}. It can 
 also be viewed as a parallel transport of the electric field $G_{i0}(\vec n)$ Eq(\ref{ef}) to $\infty$ at fixed $T_3$  $G_{i0}(\vec n) \to \Phi_{i0}$
 \begin{equation}
 \Phi_{i0}(\vec n) =  {V_C}^{\dagger} (\vec n, \infty) G_{i0}(\vec n)  V_C(\vec n, \infty)  \label{abfi}
 \end{equation}
 The same replacement  leaves the plaquette $P_{i 0}(\vec n)$ unchanged. $P_{i0}(\vec n)=\frac{1}{N}Tr[G_{i0}]=
 \frac{1}{N}Tr[\Phi_{i0}] .$ 
 
 $\Phi_{i0}$ is a gauge-invariant field and, as a consequence of the relation $\partial_i V_C(\vec n, \infty)= -iA_i(\vec n,0)V_C(\vec n, \infty)$
 \begin{equation} 
  \partial _i \Phi_{i0}= {V_C}^{\dagger} (\vec n, \infty) D_i G_{i0}(\vec n)  V_C(\vec n, \infty) =0 \label{CCNA}
 \end{equation}
 since $D_i G_{i0}(\vec n)=0$, at least  in absence of quarks.
 
 Eq(\ref{app}) is nothing but a sum of vacuum correlation functions of gauge invariant fields $\Phi_{i0}(\vec n)$. The proper way of dealing with the parallel transports is to require that they all overlap from some point on to infinity in order to have non-zero correlation\cite{adg21}. In this way the correlation functions are nothing but  the gauge invariant correlations known in the stocastic vacuum model of $QCD$\cite{dosch} \cite{simonov} \cite{ddss} which have been numerically studied on the lattice\cite{dp}\cite{ddss}.
 
 With that definition the two dots in the second cube of fig.1 are replaced by a two-point gauge-invariant correlator and, as seen by direct calculation, the two cubes cancel exactly making the kinematic divergence zero.
 In analogy with the $U(1)$ case it can be shown that the kinematic divergence cancels to all orders of the strong coupling expansion, by use of the conservation law Eq(\ref{CCNA}).
 We have thus constructed a gauge-invariant order parameter for monopole condensation in gauge theory vacuum. The inputs are 1)basic quantum mechanics Eq(\ref{mu}),2) the idea that a monopole is legitimate if its existence does not violate gauge invariance\cite{elitzur} 3) the existence of finite correlation length in the confined phase, and scale invariance in the deconfined phase. A direct test by numerical simulations of the order parameter itself on lattice  would definitely set the problem.
 \section{Discussion}
  The main point of this work is that the monopoles condensing in $QCD$  vacuum to confine quarks 
  must live in the global group, and not in local gauge representations.
  As discussed in Sect.2 the correlation functions sensitive to  deconfinement are those of the fields $S_i(\vec n)\Im Q_{i0}(\vec n)$ with an even number of points because of charge conjugation invariance. In the spirit of the stochastic vacuum model\cite{dosch} \cite{simonov} we expect that the two point function dominates. Finite contribution to $\rho$ are irrelevant to the order parameter since by Eq(\ref{romu})
  $\langle \mu \rangle \approx \exp(- \rho)$. A finite addition to $\rho$ changes the value of $\langle \mu \rangle $ but does not affect its being zero or non zero. Only potentially divergent terms matter, or, in conclusion, the gauge invariant parallel \cite{dp} two point function. Numerical studies \cite{DMP}\cite{DMP2} show  indeed that the part of it exponentially  decreasing at large distances
  vanishes at the deconfining transition. 
  
  Finally the new construction  solves an old problem first raised in \cite{GW} and then analysed in \cite{dmo}. Flux tubes joining confined quarks should keep memory of the orientation in colour space of the condensing monopoles. Flux lines should have a non trivial distribution of their orientation in colour space. In lattice simulations instead the distribution is uniform. This was indicated in \cite{GW} as a difficulty of the mechanism of dual superconductivity in explaining confinement. If instead the orientation of monopoles is fixed at infinite distance any memory of it is lost in the parallel transport from infinity.


\begin{thebibliography}{999}
\bibitem{adg21} A.~Di Giacomo, JHEP {\bf 02} 208 (2021)
 \bibitem{D} A.~ Di Giacomo, Nucl.Phys{\bf A 702}, 73 (2002)
 \bibitem{'tHP} G.~'t~Hooft, in \textit{High Energy Physics: Proceedings EPS International Conference, Palermo,Italy 23-28 june 1975, A. Zichichi ed.}(Ed. Compositori, Bologna, 1976) 1225
 \bibitem{m} S.~Mandelstam, Phys. Rep. {\bf 23C}, 245 (1976).
   \bibitem{Dig}
 A.~Di Giacomo, Acta Phys. Polon. {\bf  B 25}, 215 (1994) 
 \bibitem{DP}
A.~Di Giacomo, G.~Paffuti, Phys. Rev. {\bf D56}, 6816 (1997)
 \bibitem{ddpp} L.~Del~Debbio,~A.~Di~Giacomo,~G.~Paffuti,~P.~Pieri, Phys.Lett. {\bf B355}, 255 (1995)
 \bibitem{Dirac} P.~.A.~ M.~ Dirac, Phys.Rev {\bf 74} 817 (1948)
 \bibitem{tH} G.~'t Hooft, Nucl.Phys{\bf B79} 276 (1974)
\bibitem{Poly}A.~M.~Polyakov, JETP Lett.{\bf 20} 194 (1974)
\bibitem{'tHoo} G.~'t~Hooft, Nucl.Phys.{\bf B190} , 455  (1981)
  \bibitem{elitzur} S.~ Elitzur, Phys. Rev. {\bf D12} 3978 (1975)
  \bibitem{dlmp} A.~ Di Giacomo,~B.~Lucini,L.~Montesi, G.~Paffuti, Phys.Rev.  {\bf D 61} 034503 (2000)
   \bibitem{ddd} G.Cossu, M.~D'Elia, A.~Di~Giacomo, B.~Lucini,C.~Pica PoS Lattice 2007 296 (2007)
 \bibitem{bcdd} C.~Bonati, G.~Cossu, M.~D'Elia, A.~Di Giacomo, Phys. Rev. {\bf D85} 065001 (2012)
 \bibitem{FM} J.~ Frolich, P.~A.~Marchetti, Commun. Math. Phys.{\bf 112}, 343  (1987)
\bibitem{PC} G.~Paffuti,V.~Cirigliano, Commun,Math.Phys. {\bf 200} 381  (1999)
\bibitem{dosch} H.~G.~Dosch, Phys.Lett {\bf B190} 177 (1987)
 \bibitem{simonov} Yu.~A.~Simonov, Nucl.Phys. {\bf B 307} 512 (1988)
  \bibitem{ddss} A.~Di Giacomo,H.~G.~Dosch,V.~I.~Shevchenko, Yu.~A.~Simonov, Phys. Rept. {\bf 372} 319 (2002)
 \bibitem{dp} A.~Di Giacomo, H.~Panagopoulos, Phys.Lett {\bf B285} 133 (1992)
  \bibitem{DMP} A.~Di Giacomo, E.~ Meggiolaro, H.~Panagopoulos, Nucl.Phys.{\bf B 483} 371 (1997)
 \bibitem{DMP2} A.~Di Giacomo,~ E.~ Meggiolaro,~ H.~Panagopoulos, Nucl.Phys.{\bf B 54A} (Proc.Suppl.) 343 (1997)
 \bibitem{GW} J.~Greensite, J.~Winchester, Phys.Rev {\bf D 40} 4167 (1989)
 \bibitem{dmo}A.~Di Giacomo, M.~Maggiore, S.~Olejnik, Nucl.Phys.{\bf B347}, 441 (1990)
 \end{thebibliography}
\end{document}